\documentclass[aps,prl,twocolumn,showpacs]{revtex4}
\usepackage{graphicx}
\usepackage{times}
\usepackage{latexsym}
\def\beq{\begin{eqnarray}}
\def\eeq{\end{eqnarray}}
\def\bey{\begin{eqnarray}}
\def\eey{\end{eqnarray}}

\newcommand{\lrf}[2]{\left(\frac{#1}{#2}\right)}
\newcommand{\nnmb}{\nonumber}

\def\lsim{\mathrel{\raise.3ex\hbox{$<$\kern-.75em\lower1ex\hbox{$\sim$}}}}
\def\gsim{\mathrel{\raise.3ex\hbox{$>$\kern-.75em\lower1ex\hbox{$\sim$}}}}

\newcommand{\be}{\begin{equation}}
\newcommand{\ee}{\end{equation}}

\newcommand{\gev}{{\rm ~GeV }}

\begin{document}

\title{Multi-Component Dark Matter}  
\author{Kathryn M. Zurek$^{1,2}$}
\address{$^1$Particle Astrophysics Center, Fermi National Accelerator Laboratory, Batavia, IL  60510 \\
$^2$Department of Physics, University of Michigan, Ann Arbor, MI 48109 }

\date{\today}

\begin{abstract}

We explore multi-component dark matter models where the dark sector consists of multiple stable states with different mass scales, and dark forces coupling these states further enrich the dynamics.  The multi-component nature of the dark matter naturally arises in supersymmetric models, where both $R$ parity and an additional symmetry, such as a $Z_2$, is preserved. We focus on a particular model where the heavier component of dark matter carries lepton number and annihilates mostly to leptons.  The heavier component, which is essentially a sterile neutrino, naturally explains the PAMELA and synchrotron signals, without an excess in antiprotons which typically mars other models of weak scale dark matter.  The lighter component, which may have a mass from a GeV to a TeV, may explain the DAMA signal, and may be visible in low threshold runs of CDMS and XENON, which search for light dark matter.

\end{abstract}
%\preprint{FERMILAB-PUB-07-587-A}
\maketitle

%\section{Introduction}

There have been many tantalizing signals which may be evidence for particle dark matter.  Most recently, the PAMELA experiment has reported an cosmic ray positron excess of positrons with energy in the 10-100 GeV range \cite{Pamela}, which is consistent with annihilating dark matter \cite{theoryexcess}, confirming the excess observed by the HEAT  \cite{Heat} and AMS \cite{AMS} experiments.  The ATIC and PPB-BETS balloon experiments have likewise observed an excess, consistent with the PAMELA, HEAT and AMS results.  ATIC and PPB-BETS suggest a dark matter particle annihilating to leptons with mass in the 500-800 GeV range \cite{ATIC}; the other observations are consistent with dark matter mass in this range.  The recent Fermi results suggest, however, that the ATIC excess may be instrumental in origin.  If this is the case, the annihilating DM particle may be much lighter, with mass in the $\sim 100-200 \mbox{ GeV}$ range to explain the PAMELA excess only.  In addition, there is the observation of the synchrotron radiation toward the galactic center, the so-called ``WMAP haze,'' which is indicative of dark matter annihilating to electrons which emit photons in the galactic magnetic field \cite{Haze}.  Indeed, an annihilation cross-section to $e^+e^-$ which produces the WMAP haze is roughly the right size (up to a boost factor) to produce the AMS, HEAT, ATIC, PPB-BETS and PAMELA excesses.  The size of these signals is also roughly consistent with the freeze-out annihilation cross-section predicted for a thermal relic WIMP.  
%The EGRET experiment may also have an excess in photons, which is being further examined by the Fermi satellite \cite{egret}.  
In direct detection, the DAMA  experiment has reported an $8.2\sigma$ significance modulation in the rate of recoils in their experiment \cite{Bernabei}.  The phase and amplitude of their signal is consistent with a light elastically scattering WIMP with mass in the $\sim3-10$ GeV range \cite{PetrielloZurek} (though see \cite{DAMAfits} for a discussion of the effect of the lowest DAMA recoil bin on the fit in this window).

While these signals are intriguing, detailed explanations of these signals in terms of standard models of WIMP dark matter, such as supersymmetry, may be challenging.    
One difficulty in explaining the AMS, HEAT, PAMELA, ATIC, PPB-BETS and haze excesses is that the dark matter must have a large annihilation cross-section to leptons and a small annihilation cross-section to hadrons, since the data shows a positron excess but no excess of anti-protons \cite{Pamela,HooperWeiner,Cirelli}.  This is challenging for two reasons.  First, hadrons carry an enhancement in the annihilation cross-section which goes like $N_c$, the number of colors; hence in many models, annihilation to colored particles is the preferred mode.  Secondly, when the dark matter particle is Majorana, as in SUSY models, there is a chiral suppression which disfavors annihilation to light modes.   In SUSY, annihilation to $\bar{b}b$, $\tau^+\tau^-$, and $W^+W^-$ is preferred; it has been shown that an annihilation cross-section big enough to produce the positron excess through this mode will produce too many anti-protons through the hadronic decays of these states (see e.g. \cite{Cirelli,Salati} for the case of $W^+W^-$).

In this paper we develop models which naturally overcome this challenge,  where the dark matter effectively carries lepton number, and hence annihilation to leptons is the only mode allowed.   %The Lagrangian for the class of models we consider here was first discussed in \cite{ADM}, in the context of Asymmetric Dark Matter, where the dark matter density was set by.  
We also show that within this class of models, the dark matter may in fact also quite naturally be multi-component.  A heavier component explains the PAMELA and synchrotron excesses, while the lighter component, residing in the hidden sector, may have a much lower mass, and may explain the 
DAMA signal.  %In certain of the models we discuss here, the lighter component may have its mass set by the baryon asymmetry, rather than thermal freezeout, as discussed in \cite{ADM}.  In this case, these models predict the mass of the lighter component of dark matter to be in the few GeV range, as required by DAMA.  
These low mass states may be reachable with low threshold analyses currently being planned by the CDMS and XENON experiments \cite{lowthresh}.   

The addition of these low mass hidden sectors with multi-component dark matter naturally suggests rich dynamics in the hidden sector.  In many cases, there are new forces, both scalar and vector, which give rise to novel phenomenology, and in many ways, the rich dynamics of these low mass hidden sector dark matter models is motivated by the Hidden Valley \cite{hv}.  The components of the model we discuss here, with multiple dark forces and low mass dark matter states coupled to the SM through kinetic mixing or TeV mass states, resemble features of the low mass hidden dark matter models constructed in \cite{ADM,Pospelov,HooperZurek}.
%For example, the mediator between the hidden sector leptonic dark matter and the standard model may also lead to a Sommerfeld enhancement of the annihilation cross-section.   
Because multiple forces may reside in the hidden sector, these models may also provide a natural context for solving a second challenge for a model of DM explaining the positron excesses.
That is, there must be a boost in the annihilation of the dark matter in the halo today relative to the cross-section required at thermal freeze-out, $\langle \sigma v \rangle \sim 3 \times 10^{-26} \mbox{ cm}^3/\mbox{s}$.   For dark matter in the 500-800 GeV range, the boost is typically quite large, $\sim 100-1000$, for direct annihilation to $e^+e^-$ \cite{HooperWeiner,ATIC}.  A smaller, though still significant, boost is required for lighter DM in the $\sim 100 \mbox{ GeV}$ range \cite{HooperWeiner}.  The boost factor may come from a large overdensity in the dark matter locally in the galaxy, though simulations suggest that a boost factor much larger than $\sim5$ is difficult to produce.  A boost factor may instead imply that the size of the dark matter annihilation cross-section in the halo today is larger than the annihilation at thermal freeze-out.
%, since the annihilation rate depends on the density times the annihilation cross-section, so that instead of boosting the density, one instead boosts the annihilation cross-section.   
A possible source of the needed enhancement of the cross-section today is the so-called Sommerfeld effect \cite{Nojiri}.  This effect gives rise to an enhancement of the annihilation cross-section at low velocity $v$, so that the annihilation cross-section for particles locally in our halo ($v \sim 10^{-3}$) is enhanced with respect to the freeze-out cross-section ($v \sim 0.3$).  One of the additional dark forces may provide for such an enhancement. (See \cite{nontherm} for a model where late decay of a meta-stable state produces the needed boost.)

Models with such large boosted annihilation cross-sections potentially run into phenomenological constraints.  It was shown in \cite{Cirelli} that gamma ray constraints from HESS in the galactic center and galactic ridge require $ B \langle \sigma v \rangle \lesssim 10^{-24} \mbox{ cm}^3/\mbox{s}$, where $B$ is the astrophysical boost factor, for DM in the several hundred GeV mass window.  This constraint is problematic for the window preferred by ATIC for larger DM masses.  For smaller DM masses consistent with the PAMELA signal the constraint is less significant due to the smaller annihilation cross-sections required.  The constraint can be alleviated, however, even for dark matter in the several hundred GeV range.   The astrophysical boost factor may be smaller in the galactic center than it is at the solar radius, where the positrons originate.  This is expected where tidal disruption of dense objects in the galactic center occurs.  Thus while  $ B \langle \sigma v \rangle \lesssim 10^{-24} \mbox{ cm}^3/\mbox{s}$ at the galactic center, $B \langle \sigma v \rangle$ may be much larger locally to produce the large positron excesses.  A second less stringent constraint on these Sommerfeld boosted cross-sections is derived from BBN \cite{BBN}, $\sigma v \lesssim 7 \times 10^{-24} \mbox{ cm}^3/s$ for annihilation to electrons and $\sigma v \lesssim 2 \times 10^{-23} \mbox{ cm}^3/s$ for annihilation to muons and taus.  These constraints can be satisfied even for moderately large DM masses in the class of models we consider with boost factors which are on the order of a few.

Although dark matter with multiple components could potentially be quite complicated, in this paper we propose that the dark sectors take on a simple basic structure.  (See \cite{HyeSung} for another model of dark matter with more than one stable state.) To the SM sector, we add an ``$X$ sector.''  The $X$ sector contains generally high mass states, in the 100's of GeV range, which communicate to the SM through ${\cal O}(1)$ operators which carry lepton number.  Thus the lightest state in this sector annihilates primarily to electron-positron (or unobserved neutrino) pairs, producing the observed PAMELA and synchrotron excesses, without any excess in anti-protons.  The dark matter is essentially a sterile neutrino, which is stable by virtue of a $Z_2$ symmetry.  

When such a model is supersymmetrized, there is now a second stable state by a $R$-parity.  To the $X$-sector, we may add a hidden dark matter (hDM) sector.   The hDM sector does two things.  First, the hDM sector makes the MSSM LSP unstable to decay to  hDM states with the same $R$-symmetry charge.  Thus the lightest $R$-symmetry odd state may in fact be much lighter than the weak scale.   This was pointed out in \cite{SUSYHV} for hidden valleys and applied to dark matter in the context of a supersymmetric MeV hidden sector in \cite{HooperZurek}.   Here we are interested in the case where that state has a mass in the 1-10 GeV range, and explains the DAMA signal.  Second, the hDM sector provides a means for breaking the symmetry of the new dark forces and giving masses to the gauged mediators.  In some cases, these forces may give rise to a Sommerfeld enhancement.  The mass of those dark forces should be in the sub-50 GeV range, since the Sommerfeld enhancement is effective for mediator masses $m_M$ satisfying $g_D^2 M_X/(4 \pi) \gtrsim m_M$, where $g_D$ is the coupling of the dark force to the dark matter $X$.  Thus for $g_D \sim 0.1-1$ we see that the 1-10 GeV range is motivated both for mediators of the Sommerfeld effect and for a dark matter candidate to explain the DAMA signal.  

Such light scalar or gauged mediators are natural in the presence of hidden sectors, as shown in \cite{HooperZurek} in the context of MeV dark matter.   On the basis of naturalness considerations, one expects scalar forces (or massive gauged particles which get their masses from such scalars) to be at the weak scale.  However, such light scalars can be natural if the hidden sector is shielded from MSSM SUSY breaking (which tends to push the mass of the force mediators to the weak scale) by a weak coupling to the MSSM sector.  We will consider the case where the weak coupling is either a mixing angle $\theta$ between the dark force $U(1)_D$ and hypercharge $U(1)_Y$, or a small coupling $\lambda_D$ of a visible sector singlet scalar with the hDM sector.  These weak couplings set the mass scale $m_D$ in the hDM sector to be $m_D \sim \theta m_{SUSY}$, or $m_D \sim \lambda_D m_{SUSY}$ (up to loop factors)  where $m_{SUSY} \sim 0.1-1 \mbox{ TeV}$ are the MSSM SUSY breaking masses and $m_D$ is the typical scale for the dark forces and the DM in the hidden sector.  Since the kinetic mixing between the two sectors may typically be a loop factor $\theta \sim 10^{-2}$, or a somewhat small coupling $\lambda_D \sim 10^{-1-2}$, the low mass 1-10 GeV scale is further motivated.   While such a mechanism was introduced in the context of MeV dark matter for smaller mixings $\theta \sim 10^{-5}$, it was shown to be quite general for higher mass hidden sectors in the 0.1-100 GeV range \cite{FengKumar,ArkaniWeiner}.

%\section{A Model for the $X$ Sector:  Leptonic Dark Matter}

We now turn to constructing the $X$ model explicitly.   To the MSSM we add
\be
\Delta W = y'_i L_i H' \bar{X} + \lambda_X S_X \bar{X} X + \kappa_X S_X^3 
\label{SUSYLDMW}
\ee
where $H'$ is an electroweak doublet. There is a $Z_2$ symmetry under which $H'$ and $X$ are odd, and also an $R$-symmetry.  If a component of $X$ is the lightest $Z_2$ odd particle, it is a stable dark matter candidate, and it effectively carries lepton number, explaining why it annihilates predominantly to leptons (for another leptophilic model see \cite{ErichPaddy}) through $t$-channel $H'$ exchange.    The mass of such a dark matter state is in the 100's of GeV range, and it must be fermion to get an $s$-wave annihilation cross-section (which is unsuppressed at low velocities in the halo today).
The annihilation cross-section is
$
\sigma_{\rm ann} v = {y'_i}^4 m_{X}^2 /(16 \pi m_{H'}^4),
$
which must be $\sigma_{\rm ann} v \simeq 3 \times10^{-26} \mbox{cm}^3/\mbox{s}$ in order to be consistent with the observed relic abundance.  Thus for $m_{X} \approx 700 \mbox{ GeV}$, we find $y'_i \lesssim 0.6 $.  The scalar component $\tilde{X}$ annihilating through $t$-channel Higgsino $\tilde{H'}$ exchange to $e^+ e^-$, on the other hand, gives a $p$-wave suppressed annihilation.  
Thus to have a viable model, $\tilde{X}$ must be heavier than $X$, and rapidly decay to the $X$ fermion plus the lightest $R$-symmetry odd state, which will reside in the hDM sector.   With this annihilation cross-section to electron-positron pairs, the rate is a factor $\sim 100-1000$ below what is required to reproduce the ATIC and PAMELA signals together, and a factor $\sim 10$ below what is required to produce the PAMELA signal alone.  
The required boost from a Sommerfeld enhancement may be mediated  by a singlet scalar $S_X$ which generates the mass for $X$, $\lambda_X \langle S_X \rangle = m_X$.  This enhancement is relevant if  $\lambda_X^2/(4 \pi) m_{X} \gtrsim m_{S_X}$ is satisfied.   Since $\langle S_X \rangle = \frac{m_{S_X}}{3 \kappa_X}$, we find that the Sommerfeld condition is satisfied if $\frac{\lambda_X^3}{12 \pi} \gtrsim \kappa_X$, which is fulfilled for $\lambda_X \simeq 1$ and a relatively small $\kappa_X$.  This singlet $S_X$ must have a relatively small mixing angle with the Higgs in order not to violate direct detection bounds for the DM candidate $X$ (though it may be possible that this scalar is that of the NMSSM, see \cite{NomuraThaler} for a possible model).  We will see next that one of the scalars residing in the hDM sector may also quite naturally mediate the boost.

To this point, we have two stable states: the DM fermion $X$ stable by the $Z_2$ and the LSP (either the scalar $\tilde{X}$ or an MSSM superpartner).  With the addition of a supersymmetrized low mass hidden sector, the LSP becomes unstable to decay to the hidden sector, so that the LSP mass may be much lighter than the weak scale.  
For the purposes of this toy model, we consider the minimal hDM superpotential,
\be
W_h = \lambda_D S_D \bar{D} D + \kappa_D S_D^3.
\label{hiddenW}
\ee
This hidden toy model is fashioned after that discussed in \cite{HooperZurek}, and is to be added to the $X$-sector super-potential, Eq.~(\ref{SUSYLDMW}).  Here $S_D$ is a dark singlet field, and the dark Higgses, $\bar{D},~D$, may be charged under a new hidden gauge group $U(1)_D$, which is a dark force.  $U(1)_D$ mixes with hypercharge through the kinetic term  $\theta F_D^{\mu \nu}F_{\mu \nu}$.

The lightest state in this hDM sector may be a candidate to explain the DAMA signal, if its mass is in the 1-10 GeV range.  This mass may naturally be induced radiatively from two sources.  
First, kinetic mixing between hypercharge and $U(1)_D$ is $\theta \sim 10^{-2}-10^{-3}$, as expected when the mixing is induced by a loop of heavy particles \cite{Holdom}.  This kinetic mixing introduces SUSY breaking into the hidden sector by a two loop gauge mediation diagram, with messengers in the loop, as in \cite{HooperZurek}.  We term this mechanism for SUSY breaking in the hidden sector ``little gauge mediation.''  The size of the radiatively induced $D,~\bar{D}$ masses is
$
m_{D,rad}^2 =  \frac{3}{5} g_{D}^2 g_{Y}^2 \theta^2 m_{SUSY}^2, 
%\label{positiveMsq}
$
where $m_{SUSY} = \langle F_{mess} \rangle/(16\pi^2M_{mess})$ is the SUSY breaking mass in the messenger sector, $g_D$ is the gauge coupling of $U(1)_D$ and $g_Y$ the hypercharge gauge coupling.  With $\theta \sim 10^{-2}-10^{-3}$, and ${\cal O}(1)$ couplings, we can see that the GeV mass scale is naturally generated in the hidden sector.  In order to break $U(1)_D$, this mass-squared must be negative.  One loop graphs with the scalar $S_D$ in the loop may easily induce such a negative mass-squared,
$
m_{D,rad}^2 \simeq -\frac{4 \lambda_D^4 m_{S_D}^2}{16 \pi^2} \log \left(\frac{\Lambda^2}{m_{SUSY}^2} \right),
$
where $\Lambda$ is the scale where the soft masses are generated, and $m_{S_D}^2$ is the soft SUSY breaking mass of $S_D$ (we assume that the singlet receives a moderate SUSY breaking mass in the 10 to 100's of GeV range through a coupling to the SUSY breaking messenger fields).  For $\lambda_D \approx 10^{-1}-1$, soft masses for $D$ in the few GeV range result which are negative, even with the contribution from little gauge mediation included.

With $\langle S_D \rangle = 0$ and $\langle D, \bar{D} \rangle \neq 0$, we review the spectrum briefly, but see \cite{HooperZurek} for details.  With ${\cal O}(10^{-1-2})$ gauge coupling $g_D$ and Yukawa term $\lambda_D$, all masses in the hidden sector are ${\cal O}(\mbox{GeV})$.  The $U(1)_D$ symmetry is broken by $\langle D, \bar{D} \rangle$ and the gauge boson acquires a mass.  We have scalar mass eigenstates $m_{D_1}^2 = -  \frac{4 g_D^2-2 \lambda_D^2}{\lambda_D^2} m_{D,rad}^2$, $m_{D_2}^2 = -2 m_{D,rad}^2$, and $m_{U_D}^2 = 4 g_D^2 \langle D \rangle^2$ from the breaking of the $U(1)_D$ with $\langle D,\bar{D} \rangle^2=-m_{D,rad}^2/\lambda_D^2$.   The fermion masses arise through $\tilde{D},\tilde{\bar{D}},\tilde{U}_D,\tilde{S}_D$ mixing, two with masses $2 g_D \langle D \rangle$ (a $\tilde{U}_D$ gaugino-$\tilde{D}$ Higgsino mix) and two with masses $\sqrt{2}\lambda_D \langle D \rangle$ (a $\tilde{S}_D$ singlino-$\tilde{D}$ Higgsino mix).  We assume $g_D \gtrsim \lambda_D$ so that the fermions with mass $\sqrt{2} \lambda_D \langle D \rangle$ are stable dark matter candidates, provided they are lighter than the gravitino. 

Now, we can see that such a sector can plausibly give rise to a signal in DAMA in the elastically scattering WIMP window.  We take the $\tilde{D}-\tilde{S}_D$ fermions to be the dark matter with mass $m_{hDM}$ in the 3-10 GeV range.   
The DM may annihilate to the axion associated with the angular components of $D,\bar{D}$ which is light, as in the NMSSM. The annihilation cross-section of the hidden dark matter to these axions is 
\beq
\sigma_{ann}  &\simeq&  \frac{\lambda_D^4}{16 \pi} \frac{1}{m_{hDM}^2}  \\
&\sim&  10^{-35} \mbox{ cm}^2 \left(\frac{\lambda_D}{0.1}\right)^2 \left(\frac{8~ {\rm GeV}}{m_{hDM}}\right)^2.  \nonumber
\eeq
This cross-section is of the order $\sim 10^{-36} \mbox{ cm}^2$ necessary to produce the correct relic density (this candidate need not be all the dark matter). 
The direct detection cross-section by exchanging a $U_D$ gauge boson is
\beq
\sigma_{SI} & \simeq &  \frac{g_D^2 g_Y^2 \theta^2}{\pi} \frac{m_{r}^2}{m_{U_D}^4} \\
& \sim & 10^{-40} \mbox{ cm}^2 \left(\frac{g_D g_Y \theta}{10^{-4}}\right)^2  \left(\frac{8 \mbox{ GeV}}{m_{U_D}} \right)^4\nonumber,
\eeq
where $m_r$ is the reduced mass of the nucleon-DM system.  We see that a hidden sector, which simultaneously generates natural GeV mediators and GeV scale dark matter candidates, produces a direct detection cross-section in a range to be the explanation for the DAMA signal.

In addition, if the singlet $S_D$ couples to the visible Higgs through a term in the superpotential $\zeta S_D H_u H_d$, this provides an additional channel for direct detection.  The size of the scattering cross-section is
\beq
\sigma_n &\simeq& \frac{m_r^2}{2 \pi}N_n^2\,
\lrf{\lambda_D\zeta\,v_u\langle D \rangle}{m_{h^0}^2}^2\frac{1}{m_{D_1}^4}\\
&\simeq& 2\times 10^{-41}\, \mbox{ cm}^2\,
\lrf{N_n}{0.1}^2\lrf{\lambda_D}{0.1}^2
\lrf{\zeta}{10^{-2}}^2\nnmb\\
&&\times\lrf{\langle D \rangle}{20\,\gev}^2\lrf{100\,\gev}{m_{h^0}}^4
\lrf{10\,\gev}{m_{D_1}}^4,\nnmb
\eeq
where $N_n$ comes from the effective coupling of the
exchanged scalar to the target nucleus and $h^0$ is the MSSM Higgs.  We see again that this mechanism results in a scattering cross-section is in the  $10^{-41}-10^{-39} \mbox{ cm}^2$ window for explaining the DAMA signal with light WIMPs (if the light state only composes a fraction of the DM, scattering cross-sections should be correspondingly larger).  Alternatively, if the DAMA signal turns out not to be from DM scattering, it is easy to evade direct detection bounds by lowering the mixing $\zeta$ or correspondingly raising the mass of the mediators; these lower mass WIMPs may still be in reach of the low threshold runs of CDMS \cite{lowthresh} and XENON.

The general conclusion here is that such hidden sectors with GeV mass dark matter particles and dark forces of GeV mass mediators arise naturally in a framework where the hidden sector communicates to the SM through kinetic mixing of dark force with hypercharge, or through mixing of a singlet scalar with both the hidden and visible sectors.  The mixing simultaneously provides motivation for observation of these states by direct detection experiments.  These light gauged or scalar mediators may in fact mediate the Sommerfeld enhancement as well, if $X$ is charged under the $U(1)_D$, or if $S_D$ also couples to $X$ in addition to $D$.  %The GeV scale is generated through radiative effects from the $S_D$ scalars, and also through the mixing, $m_{hDM} \sim \theta m_{SUSY}$.  The mixing simultaneously provides motivation for observation of these states by direct detection experiments.

%\section{Conclusions}

We have discussed multi-component dark matter models in which the dark sector is more complex than a single weakly interacting field.  In many cases, these models give rise to additional dark forces which enrich the dark matter dynamics.   Phenomenologically, the focus of this paper has been on explanations of the PAMELA, ATIC, PPB-BETS, HEAT, AMS, and DAMA excesses.  In the models discussed here, the dark matter candidate which explains the positron excess carries lepton number; it is stable by an additional $Z_2$ symmetry.  We showed that in supersymmetric models of this type, there are naturally two dark matter candidates--the lighter candidate may explain the DAMA signal, and may be observable by low threshold runs of CDMS, XENON.  We also showed how dark forces that arise in hidden sector dark matter models may naturally have their masses generated at the GeV scale, further motivating the low mass WIMP window as a well-motivated scale for direct detection of dark matter.  Dark matter dynamics and dark matter sectors may be rich.  As multiple experiments with varied detection techniques probe the dark sector, we may discover a dark hidden world in lieu of a single weakly interacting particle.

%In particular, relevant for indirect detection experiments, the correlation between the annihilation

\bigskip

This work has been supported by the US Department of Energy, including grant DE-FG02-95ER40896, and by NASA grant NAG5-10842.  We thank Paddy Fox, Dan Hooper, Peter Ouyang, Frank Petriello, and Erich Poppitz for discussions, and Alessandro Strumia for a comment on the first version.

\end{document}